\documentstyle[preprint,tighten,pra,aps,amssymb,exscale]{revtex}
\begin{document} \draft
\newcommand{\mythbb}[1]{{\Bbb #1}}
\newcommand{\mythfrak}[1]{{\frak #1}}
\newcommand{\mythbf}[1]{{\bf #1}}
\newcommand{\mythcal}[1]{{\cal #1}}
\newcommand{\C}{\mythbb C}
\newcommand{\R}{\mythbb R}
\newcommand{\Z}{\mythbb Z}
\renewcommand{\k}{\Bbbk}
\newcommand{\A}{\mythcal A}
\newcommand{\D}{\mythcal D}
\newcommand{\T}{\mythcal T}
\newcommand{\I}{\mythbf{1}}
\renewcommand{\O}{\mythbf{0}}
\newcommand{\Matr}[2]{\left( \begin{array}{#1} #2 \end{array} \right)}
\newcommand{\eq}[1]{Eq.~(\ref{#1})}
\def\eqtag{\eqnum}
\newcommand{\PA}{\mythcal S}
\def\astar{\overstar}
\newcommand\fG{\mythfrak}
\newcommand\nA{\mythbf{\breve{\fG A}}}
\def\nplus{\smallsmile}
\newcommand\Ginv[1]{{\mythbf{\dot{\mit #1}}}}

\title{\LARGE \bf
 Symmetries of complexified space-time and algebraic structures
 (in quantum theory).
}
\author{Alexander Yu.\ Vlasov\thanks{E-mail:
{\tt Alexander.Vlasov@pobox.spbu.ru}}}
\address{FRC/IRH, Mira Street 8, St.-Petersburg, Russia}

\maketitle

\sloppy

\begin{abstract}
In the paper is discussed description of some algebraic structures in  quantum
theory by using formal recursive constructions with ``complex  Poincar\'e
group'' $ISO(4,\C)$.
\end{abstract}

\section{Introduction}

At 1989, on the occasion of celebration of famous E. Wigner's 1939 paper
\cite{wig}, S.~Weinberg wrote \cite{wei}:
\begin{quotation}
\small
\noindent\ldots\ I want here to take up a challenge in footnote {\bf 1} of
Wigner's 1939 paper. The first sentence of Wigner's paper is as follows:

``{\bf It is perhaps the most fundamental principal of quantum mechanics that the
system of states forms a linear manifold.$\bf\!^1$\ldots}''\\
Looking down at the bottom of the page, one finds that footnote {\bf 1} says:

``{\bf The possibility of a future non-linear character of the quantum
mechanics must be admitted of course.}''

In the succeeding half-century little has been done with the idea that
quantum mechanics may have small non-linearities. \ldots
\end{quotation}

The {\em fundamental principal} mentioned above by
E.~Wigner may be considered as yet another example of {\em miracle
effectiveness of algebraic structures in quantum mechanics}, because it
is related with possibility of representation of a group as a subset of
an {\em algebra} of linear operators on the linear manifold of states. In
such approach {\em linearity} is consequence of {\em distributivity}
of algebraic operations ``$+$'' and ``$\cdot$'': $a(b+c) = ab + ac$.

On the other hand, the Wigner's construction of representations of Poincar\'e
group developed in \cite{wig} may be considered as an example of {\em the
theory of induced representation of semidirect products} \cite{ThGr}, i.e.
the existence of a linear manifold here is a part of a structure of representation
of the particular class of groups with an abelian subgroup (translations).

{\em So maybe success of algebraic approach in quantum mechanics is close
related with a structure of the symmetry group of space-time, or otherwise,
the structure of space-time follows from basic principles of quantum theory?}

It is not possible to answer such a difficult question in a short article
and here is simply discussed some formal structures with groups, rings
and algebras, possibly relevant with such kind of questions.

\section{From algebras to groups and back again}
\subsection{From algebras to groups ... }

In this paper are not discussed Lie algebras of Lie groups and instead of
it are used more simple structures like group of invertible elements of
given algebra, but even such approach produces some interesting construction.
Let us accept view on an
algebra as {\em a group with operators} \cite{vdw}. More generally, we
can consider the algebra even as {\em a set $\A$ with four kinds of operators}:
\begin{equation}
\begin{array}{lrll}
 {\sf S}_b &:& a \mapsto a + b;& a,b \in \A, \\
 {\sf L}_l &:& a \mapsto l a;& a,l \in \A, \\
 {\sf R}_r &:& a \mapsto a r;& a,r \in \A, \\
 {\sf C}_\lambda &:& a \mapsto \lambda a;& a \in \A, \lambda \in \k.
\end{array}
\label{algop}
\end{equation}
where $\k$ is field of coefficients of the algebra $\A$, for example
real or complex numbers, $\R$, $\C$, but the last family of operator
is written here for completeness and is not discussed\footnote{%
I.e. here we are considering only structure of {\em ring} $\A$.}
in detail because of some difficulties mentioned below.

For an associative algebra $\A$ all operators \eq{algop} --- are
transformations and if to consider only invertible $l,r$ (and
$\lambda \ne 0$), then all possible compositions of the operators
${\sf S}_b, {\sf L}_l, {\sf R}_r$ (and ${\sf C}_\lambda$) --- are some
group of transformations\footnote{For arbitrary $l,r$ it is {\em semigroup}.}
of set $\A$. Let us denote $\Ginv\A$ group of invertible elements of
algebra $\A$, i.e. for $l,r \in \Ginv\A$, $\exists l^{-1},r^{-1} \in \Ginv\A$.

\medskip

Let us consider first structure of group $\D(\A)$ of transformations generated
by elements ${\sf S}_b, {\sf L}_l$; $(b \in \A, l \in \Ginv \A)$ --- the group
is simpler, but encapsulate full structure of the algebra $\A$, i.e. laws of
addition and multiplication.
If the algebra $\A$ may be represented as subset of algebra of $n{\times}n$
matrices any element of $\D(\A)$ may be represented as $(2n{\times}2n)$
matrix. It is enough to represent elements of $a \in \A$ as $(n{\times}2n)$
matrix:
\begin{equation}
  a \longleftrightarrow \Matr{c}{a \\ \I},
\label{an2n}
\end{equation}
then an operator ${\sf S}_b$ may be written as $(2n) \times (2n)$ matrix:
\begin{equation}
  {\sf S}_b \longleftrightarrow \Matr{cc}{\I & b \\ \O & \I};
\quad \Matr{cc}{\I & b \\ \O & \I}\Matr{cc}{a \\ \I} = \Matr{cc}{a + b \\ \I},
\end{equation}
and operator ${\sf L}_l$ may be written as $(2n) \times (2n)$ matrix
\begin{equation}
  {\sf L}_l \longleftrightarrow \Matr{cc}{l & \O \\ \O & \I};
\quad \Matr{cc}{l & \O \\ \O & \I}\Matr{cc}{a \\ \I} = \Matr{cc}{la \\ \I},
\end{equation}
where $\O$ and $\I$ are $n \times n$ matrix representations of zero and
unit of the algebra $\A$. It is clear, that compositions of any number of
such matrices ${\sf S}_{b_i}, {\sf L}_{l_j}$ is $(2n) \times (2n)$ matrix:
\begin{equation}
  {\sf D}_{(B,L)} \equiv \Matr{cc}{L & B \\ \O & \I};
\quad L \in \Ginv\A,\ B \in \A.
\label{SPair}
\end{equation}
More precisely:
\begin{itemize}
\item[(A)] Operators ${\sf S}_{b}, {\sf L}_{l}$ --- are simply two
 special cases of the elements \eq{SPair}:
 \begin{equation}
 {\sf S}_{b} = {\sf D}_{(b,\I)}; \quad {\sf L}_{l} = {\sf D}_{(\O,l)}.
 \end{equation}
\item[(B)] A composition of two matrices \eq{SPair} again has the
 same structure:
\begin{equation}
  \Matr{cc}{L_1 & B_1 \\ \O & \I} \Matr{cc}{L_2 & B_2 \\ \O & \I} =
  \Matr{cc}{L_1 L_2 & L_1 B_2 + B_1 \\ \O & \I},
\end{equation}
i.e.,
\begin{equation}
  {\sf D}_{(L_1 B_2 + B_1,L_1 L_2)} = {\sf D}_{(B_1,L_1)} {\sf D}_{(B_2,L_2)}.
\label{SPairComp}
\end{equation}
\item[(C)] And, finally, any element \eq{SPair} itself is simply expressed
 as a product of two basic transformations:
\begin{equation}
  \Matr{cc}{\I & B \\ \O & \I} \Matr{cc}{L & \O \\ \O & \I} =
  \Matr{cc}{L & B \\ \O & \I} \Rightarrow
 {\sf D}_{(B,L)} = {\sf S}_{B} {\sf L}_{L}.
\end{equation}
\end{itemize}

It should be mentioned also, that the order of elements in last expression
does matter because they are not commute. The noncommutativity also
related with an important expression for adjoint action of ${\sf L}_L$:
\begin{equation}
  \Matr{cc}{L & \O \\ \O & \I}\Matr{cc}{\I & B \\ \O & \I}
 \Matr{cc}{L^{-1} & \O \\ \O & \I} = \Matr{cc}{\I & L B \\ \O & \I},
\label{AdMat}
\end{equation}
i.e.,
\begin{equation}
 {\sf L}_L {\sf P}_B {\sf L}_L^{-1} = {\sf P}_{L\,B}, \quad\mbox{or}\quad
 {\rm Ad}[{\sf L}_L]({\sf P}_B) = {\sf P}_{L\,B},
\label{AdLP}
\end{equation}
where ${\rm Ad}[a](b) \equiv a b a^{-1}$ is adjoint action of some
element of a group.

Here was used an identity ${\sf L}_L^{-1} = {\sf L}_{L^{-1}}$.
An expression for inversion for an arbitrary element of $\D(\A)$
is also can be simply found and checked:
\begin{equation}
\Matr{cc}{L & B \\ \O & \I} \Matr{cc}{L^{-1} & -L^{-1}B \\ \O & \I} =
\Matr{cc}{\I &\O\\ \O& \I}
\end{equation}
i.e.,
\begin{equation}
 {\sf D}_{(B,L)}^{-1} = {\sf D}_{(-L^{-1}B, L^{-1})}.
\end{equation}

\subsubsection{Semidirect products}

It is possible to describe the same things in more general way without a
matrix representation by using {\em semidirect product} of groups.
Let $\bf B$ is abelian group with additive notation:
$b_1,b_2, b_1+b_2 \in B$ and each element of other group $\bf L$
generates some transformation of $\bf B$, then semidirect product
$\mythbf B \rtimes \mythbf L$ is set of pairs $(b,l)$,
$b \in \mythbf B, l \in \mythbf L$ with composition law\footnote{In
more general case \cite{ThGr} {\bf B} is not necessary abelian group and
multiplicative notation is used: $g_1,g_2, g_1 g_2 \in B$, any element
$\Lambda \in {\bf L}$ generates an automorphism of {\bf B}:
$\Lambda(g_1 g_2) = \Lambda(g_1) \Lambda(g_2)$,
then instead of \eq{SPairAdd} in definition is used:
\begin{equation}
(g_1,\Lambda_1) (g_2,\Lambda_2) = (g_1 \Lambda_1(g_2), \Lambda_1 \Lambda_2)
\eqtag{\ref{SPairAdd}$'$}
\label{SPairMul}
\end{equation}
} \cite{ThGr}:
\begin{equation}
   (b_1,l_1) \, (b_2,l_2) = (b_1 + l_1(b_2), l_1 l_2).
\label{SPairAdd}
\end{equation}

The \eq{SPairAdd} coincides with \eq{SPairComp}, i.e. group $\D(\A)$
can be expressed as semidirect product:
\begin{equation}
 \D(\A) = \A_+ \rtimes \Ginv\A,
\end{equation}
where algebra $\A$ considered as {\em a group with operators} mentioned earlier,
i.e. $\A_+$ is the algebra $\A$ considered as an abelian group\footnote{
$\A_+$ produced from $\A$ via ``forgetting operation of multiplication''.}
with respect to addition $(a,b) \mapsto a+b$ and elements of $\Ginv\A$ are
operators i.e. transformations of $\A_+$ represented as left multiplications
$a \mapsto l\, a$; $l \in \Ginv\A \subset \A$, $a \in \A_+ \cong \A$.

It may be useful also to consider separately actions ${\sf C}_\lambda$ of
scalars and to use subgroup $\Ginv\A_\circ \equiv \Ginv\A / \k_*$,
$\Ginv\A \cong \Ginv\A_\circ \times \k_*$ there $\k_* = \k-\{0\}$
is a multiplicative group consisting of all element of field $\k$ except zero.
It is possible to construct another semidirect product with
the group $\Ginv\A_\circ$ instead of $\Ginv\A$ :
\begin{equation}
 \D_\circ(\A) \equiv \A_+ \rtimes \Ginv\A_\circ.
\label{Dc}
\end{equation}

Simple examples of semidirect products are affine groups of motions of
Euclidean and pseudo-Euclidean spaces $ISO(n)$ and $ISO(l,m)$, classical
Galilean and Poincar\'e groups, an affine group of the general relativity.
It should be mentioned, that $\D(\A)$ also can be represented as
an affine group. If $\A$ is $N$-dimensional algebra, then instead of
$(n{\times}n)$-matrix representation of $\A$ with $n^2 \ge N$ and
$(2n{\times}2n)$ matrices \eq{SPair} acting on $(n{\times}2n)$
matrix \eq{an2n}, it is possible to use an affine group of $(N+1) \times (N+1)$
matrices acting on vectors with $N+1$ components.

Really, let us consider the algebra $\A$ as a vector space $\A_+$ and use
notation $\vec{a}$ for an element of $\A_+$, considered as a vector with $N$
components. Let us now consider an action of $\Ginv\A$ via left
multiplications: $a' = l\, a$. Because it is linear transformation for any
$l$, it can be described by $N{\times}N$
matrix $\hat{l}$, $\vec{a}' = \hat{l} \vec{a}$. So instead of action of matrix
\eq{SPair} on a rectangular matrix \eq{an2n} here is an affine group with
$(N+1)\times(N+1)$ matrices:
\begin{equation}
 \Matr{cc}{\hat{l}&\vec{b}\\ \vec{0}^T&1} \Matr{cc}{\vec{a}\\1} =
\Matr{cc}{\hat{l}\vec{a}+\vec{b}\\1},
\end{equation}
where $\vec{a},\vec{b}$ are $N$-dimensional vectors (i.e. $1 \times N$
``matrices''), $\hat{l}$ is $N\times N$ matrix, $\vec{0}^T$ is
$N \times 1$ ``matrix'' (transposed vector) with $N$ zeros, and 1 is unit
(scalar).

\medskip

A classical Poincar\'e group is an example of an affine group $ISO(3,1) =
R^4 \rtimes SO(3,1)$. A spinor Poincar\'e group \cite{wig,ThGr,AQFT}
of quantum mechanics
is also may be represented as a semidirect product $\R^4 \rtimes SL(2,\C)$,
but in a less direct way, than for affine groups. For example, it is
possible to represent the additive abelian group $\R^4$ as a four-dimensional
space of Hermitian $2\times 2$ matrices $H$ using Pauli matrices $\sigma_i$:
\begin{equation}
 \vec{v} \longrightarrow H = \sum_{i=0}^3 v_i \sigma_i,
 \quad \det(H) = \|\vec{v}\|_{Mink} \equiv v_0^2 - v_1^2 - v_2^2 - v_3^2
\label{vecmat}
\end{equation}
then a transformation generated by
some element $M \in SL(2,\C)$ and used in definition of the semidirect product
is represented as:
\begin{equation}
 H \mapsto M H M^*,\quad M \in SL(2,\C),~~H = H^*
\label{MHM}
\end{equation}
The action \eq{MHM} is the same for
$M$ and $-M$, it is the well known representation of $SL(2,\C)$ with a two-fold
covering homomorphism of Lorentz group $SO(3,1)$ \cite{ThGr,AQFT}.

 A matrix representation of the spinor Poincar\'e group is discussed below.

\medskip

Let us consider now all three operators ${\sf S}_b, {\sf L}_l, {\sf R}_r$
from \eq{algop}. First two operators was used in construction of $\D(\A)$
and now it is necessary to include also ${\sf R}_r$. The operators have
property:
\begin{equation}
 {\sf R}_{r_1} {\sf R}_{r_2} a = {\sf R}_{r_1} (a r_2) = (a r_2) r_1 =
 {\sf R}_{r_2 r_1} a,
\end{equation}
i.e., the opposite order of elements in comparison with ${\sf L}_{l_1}$.
To avoid such a problem, it is possible instead of ${\sf R}_r$ to use
${\sf R}'_r \equiv {\sf R}_{r^{-1}} = {\sf R}^{-1}_r $. It should be mentioned
also, that instead of a map $r \to r^{-1}$, here could be used any
other antiisomorphism, for example, if $\A$ is $*$-algebra it may be
${\sf R}^*_r \equiv {\sf R}_{r^*}$.

Let us denote as $\T(\A)$ a group of transformations (of $\A$) generated by
arbitrary compositions of operators ${\sf S}_b, {\sf L}_l, {\sf R}'_r$,
where $l,r \in \Ginv{\A}$. Any such transformation can be expressed
as a combined action:
\begin{equation}
 {\sf T}_{(B,L,R)}\, a = {\sf S}_B {\sf L}_L {\sf R}'_R\, a = L a R^{-1} + B
\end{equation}
for some $L,R \in \Ginv{A}$, $B \in \A$.

It is clear, that the composition of two such elements is again in $\T(\A)$:
\begin{equation}
 {\sf T}_{(B_1,L_1,R_1)} {\sf T}_{(B_2,L_2,R_2)} =
 {\sf T}_{(L_1 B_2 R_1^{-1}+B_1,L_1 L_2,R_1 R_2)}
\label{TrioComp}
\end{equation}
and coincides with definition of composition for semidirect product of the
additive abelian group $\A_+ \cong \A$ and a group $\Ginv\A \times \Ginv\A$
represented as direct product of two commuting subgroups of operators
${\sf L}_l$ and ${\sf R}'_r$, where action of an element $(l,r) \in
\Ginv\A \times \Ginv\A$ on $\A_+$ is defined simply as:
$ a \mapsto l a r^{-1}$, i.e.
\begin{equation}
\bigl(b_1,(l_1,r_1)\bigr) \bigl(b_2,(l_2,r_2)\bigr) =
\bigl(\underbrace{l_1 b_2 r_1^{-1}}_{(l_1,r_1)(b_2)} + b_1,
(l_1 l_2,r_1 r_2)\bigr)
\label{DPairComp}
\end{equation}
and so the abstract set of triples with law of composition \eq{TrioComp},
\eq{DPairComp} is semidirect product:
\begin{equation}
 \tilde\T(\A) \equiv \A_+ \rtimes (\Ginv\A \times \Ginv\A).
\end{equation}

It should be mentioned, that there is some difference between $\tilde\T(\A)$
and group of transformation $\T(\A)$ of $\A$ defined above. Action of
$\tilde\T(\A)$ on $\A$ is defined as:
\begin{equation}
\tilde\T(B,L,R) \colon a \mapsto L a R^{-1} + B
\end{equation}
and so for any element of the center $c \in {\bf C}(\Ginv\A)$ (i.e.
$c \in \Ginv\A$; $c a = a c$, $\forall a \in \A$) the action of $\tilde\T(B,L,R)
\in \tilde\T(\A)$ coincides with action of $\tilde\T(B, c\,L, c\,R)$.

So $\T(\A)$ is the quotient group of
$\tilde\T(\A)$ with respect to an equivalence relation $\tilde\T(B,L,R) \sim
\tilde\T(B, c\,L, c\,R)$ discussed above. It will be shown that the spinor
Poincar\'e group used in quantum mechanics can be represented formally
as a subgroup of $\tilde\T(\PA)$ (where $\PA$ is Pauli algebra of $2{\times}2$
complex matrices).

Let us denote $\Ginv\A_{\sf c} = \Ginv\A/{\bf C}(\Ginv\A)$,
$\Ginv\A \cong \Ginv\A_{\sf c} \times {\bf C}(\Ginv\A)$, then
it is possible to write:
\begin{equation}
 \T(\A) = \A_+ \rtimes (\Ginv\A \times \Ginv\A_{\sf c}) =
 \A_+ \rtimes ({\bf C}(\Ginv\A) \times \Ginv\A_{\sf c} \times \Ginv\A_{\sf c}).
\end{equation}

Here is also justified to consider construction with $\Ginv\A_\circ$ instead
of $\Ginv\A$:
\begin{equation}
 \tilde\T_\circ(\A) \equiv
 \A_+ \rtimes (\Ginv\A_\circ \times \Ginv\A_\circ).
\end{equation}
It is also close related with consideration of field of scalars (operators
${\sf C}_\lambda$ in \eq{algop}), because in many algebras discussed in the
paper all elements of the center are simply $c = \lambda\I, \lambda \in \k$,
or $\Ginv\A_\circ = \Ginv\A_{\sf c}$.

\medskip

If an algebra $\A$ is represented via $n \times n$ matrices, there is
interesting  representation of $\tilde\T(B,L,R) \in \tilde\T(\A)$
as $(2n) \times (2n)$ matrix:
\begin{equation}
 \tilde\T(B,L,R) \longleftrightarrow \Matr{cr}{L & BR\\ \O & R}.
\label{TrioRepr}
\end{equation}
Verification of \eq{TrioComp} is straightforward:
\begin{equation}
\Matr{cr}{L_1 & B_1 R_1 \\ \O & R_1} \Matr{cr}{L_2 & B_2 R_2 \\ \O & R_2} =
\Matr{cr}{L_1 L_2 & (L_1 B_2 R_1^{-1} + B_1) R_1 R_2 \\ \O & R_1 R_2}.
\end{equation}

If $\A$ is $*$-algebra: $(ab)^* = b^* a^*$, there is an involute
automorphism of $\tilde\T(\A)$:
\begin{equation}
 *\colon \tilde\T(B,L,R) \mapsto \tilde\T(B^*,{R^*}^{-1},{L^*}^{-1}).
\label{starTA}
\end{equation}

There is an invariant subgroup of the involution
$\astar\D(\A) \subset \tilde\T(\A)$:
\begin{equation}
 \astar\D(H,G) \equiv \tilde\T(H,G,{G^*}^{-1}),\quad
 H \in \A_+, H = H^*\!,\ G \in \Ginv\A
\label{Dinv}
\end{equation}

\subsubsection{Pauli algebra}

As an example let us consider the Pauli algebra $\PA = \C(2\times 2)$ of all
$2 \times 2$ complex matrices. In such a case the group of invertible
elements is represented by matrices with nonzero determinant
$\Ginv\PA = GL(2,\C)$ and
\begin{equation}
 \tilde\T(\PA) = \C^4 \rtimes (GL(2,\C) \times GL(2,\C)).
\label{tTPA}
\end{equation}

Here is possible to show some difficulties with description of scalars.
It is clear, that any matrix $M \in GL(2,\C)$ can be represented
as product \mbox{$M = \lambda M'$} of scalar $\lambda = \det(M)$ and some
matrix with unit determinant $M' = M/\det(M)$, $M' \in SL(2,\C)$.
But such decomposition is not unique, because the matrix $-M'$ also
has unit determinant and so $M = (-\lambda) (-M')$.

Let us denote $\PA_\circ = PGL(2,\C) = GL(2,\C)/\C_* \cong SL(2,\C)/\Z_2$,
where $\C_* = \C -\{0\}$ is multiplicative group of complex
numbers without zero and $\Z_2$ is used to express a discrete group
with two elements $\Z_2 \cong \{\I,-\I\}$. Construction of the last term
also coincides with two-fold representation of the Lorentz group discussed in
\eq{MHM} and so $PGL(2,\C) \cong SO(3,1)$.

It is possible to write
\begin{equation}
 \T(\PA) = \C^4 \rtimes (\C_* \times PGL(2,\C) \times PGL(2,\C))
\label{TPA}
\end{equation}
and
\begin{equation}
 \T_\circ(\PA) = \C^4 \rtimes (PGL(2,\C) \times PGL(2,\C)).
\end{equation}
The problem with $PGL(2,\C)$ group is because of representation as a quotient
group, i.e. a set of an equivalence classes, the pairs of two matrices
$M, -M \in SL(2,\C)$.

Spinor Poincar\'e group is a subgroup of $\astar\D(\PA)$,
where in definition \eq{Dinv} is used an element
$G \in SL(2,\C) \cong \PA_\circ \times \Z_2$ .

The \eq{Dinv} for $\astar\D(\PA)$ and matrix representation
\eq{TrioRepr} of $\tilde\T(\PA)$ produce a known matrix
representation of the spinor Poincar\'e group
$\R^4 \rtimes SL(2,\C)$ \cite{AQFT}:
\begin{equation}
(H,\Lambda) \longleftrightarrow
\Matr{cr}{\Lambda & H{\Lambda^*}^{-1}\\ \O & {\Lambda^*}^{-1}}\!,
\ H \in \C(2\times 2),\ H = H^*\!,\ \Lambda \in SL(2,\C).
\label{spinPoin}
\end{equation}

\subsubsection{Symmetries of ``complex space-time''}

The action of $\T(\PA)$ group on $a \in \C^4$ is composition of
shifts $a \mapsto a + b$, scaling (dilations)
$a \mapsto \lambda a$, $\lambda \in \C$ and $SO(4,\C)$ ``rotations''
$a \mapsto L a R^{-1}$, for some $L, R \in GL(2,\C)$, $\det(L)/\det(R) = \pm1$.
The relation of the last transformation with ``complex rotations'' described
by correspondence $\C^4 \leftrightarrow \C(2 \times 2)$ represented by
analogue of \eq{vecmat} for vector $\vec{v} \in \C^4$ with complex
coefficients $v_i$.

%The last transformation is due to isomorphism:
%\begin{equation}
% SO(4,\C) \cong SO(3,1,\C) \cong (SL(2,\C) \times SL(2,\C))/\Z_2,
%\end{equation}
%because if to represent $\C^4$ as space of complex matrices
%$\C(2 \times 2)$ using \eq{vecmat} with complex $v_i$, then any
%$SO(3,1,\C)$ transformation of $\C^4$ corresponds to $H \mapsto L H R$,
%$H \in \C(2 \times 2)$, $L,R \in SL(2,\C)$, such transformations
%save $\det(H)$ that coincides with ``complex norm'' $\|\vec{v}\|_{Mink_\C}$,
%but because pairs $(L,R)$ and $(-L,-R)$ represent the same transformation
%here again is necessary to exclude equivalence relation due to $\Z_2$-symmetry
%$(L,R) \sim (-L,-R)$.

\subsection{ ... and back again}

In the previous part were discussed groups expressed via
semidirect products like $\D(\A)$ or $\T(\A)$ for some algebra $\A$.
Here is considered an ``opposite process'' of construction of a specific
algebraic structure associated with a group represented as a semidirect
product $\mythbf{G = B \rtimes L}$. Let us again do not discuss field of
scalars and describe construction of a {\em ring} first.

\subsubsection{Nonlinear quasi-ring structure}\label{part:nring}

For any group $\bf G$ it is possible to consider a space of transformation
of the group, i.e. functions $\fG f \colon \mythbf{G \to G}$, or
$\fG f \in \mythbf{G^G}$. Such transformation is called {\em endomorphism}
$\fG e \in {\rm End}(G)\subset \mythbf{G^G}$ if
$\fG e(g_1 g_2) = \fG e(g_1) \fG e(g_2)$, $\forall g_1,g_2 \in \mythbf G$.
Let us use notation $\fG{f\,h}$ for composition $\fG f(\fG h(g))$,
$g \in \mythbf G$
and also introduce an operation $\fG{f \nplus h}$:
\begin{equation}
 (\fG{f \nplus h})(g) \equiv \fG f(g) \fG h(g),\quad
 g,\,\fG f(g),\,\fG h(g) \in \mythbf G \ \ (\fG f,\fG g \in \mythbf{G^G}).
\label{npDef}
\end{equation}

The operation has the following properties:
\begin{mathletters}
\begin{equation}
 \fG{(f \nplus g)h} = \fG{(fh) \nplus (gh)}
 \quad (\forall \fG{f,g,h} \in \mythbf{G^G}).
\label{Rdistr}
\end{equation}
\indent~~{\bf Proof:}
$\fG{(f \nplus g)\bigl(h}(g)\bigr) =
\fG{f\bigl(h}(g)\bigr)\,\fG{g\bigl(h}(g)\bigr) =
\fG{(fh \nplus gh)}(g)$.

\begin{equation}
 \fG{e(f \nplus g)} = \fG{(ef) \nplus (eg)}
 \quad (\forall \fG e \in {\rm End}(G),\ \fG{f,g} \in \mythbf{G^G}).
\label{Ldistr}
\end{equation}
\indent~~{\bf Proof:}
$\fG{e(f \nplus g)}(g) = \fG e\bigl(\fG f(g) \fG g(g)\bigr) =
\fG{e\bigl(f}(g)\bigr)\,\fG{e\bigl(g}(g)\bigr) = \fG{(ef) \nplus (eg)}(g)$.

\begin{equation}
 \left.
 \begin{array}{r}
 \fG {f,g} \in {\rm End}(\mythbf G)\\
 \fG f(g_1)\,\fG g(g_2) = \fG g(g_2)\,\fG f(g_1),\ \forall g_1,g_2
 \end{array}
 \right\} \Rightarrow
 \fG{f \nplus g = g \nplus f} \in {\rm End}(\mythbf G).
\label{CommCond}
\end{equation}
\indent~~{\bf Proof:}
$(\fG{f \nplus g})(g_1 g_2) = \fG f(g_1 g_2) \fG g(g_1 g_2) =
\fG f(g_1) \fG f(g_2)\fG g(g_1) \fG g(g_2) = $\\
\indent \qquad $= \fG f(g_1) \fG g(g_1) \fG f(g_2) \fG g(g_2) =
(\fG{f \nplus g})(g_1)\,(\fG{f \nplus g})(g_2)$.
\end{mathletters}

\medskip

Let us introduce $\fG f^\nplus \in \mythbf{G^G}$:
\begin{equation}
 \fG f^\nplus(g) \equiv \bigl(\fG f(g)\bigr)^{-1},
 \quad \fG f \nplus \fG f^\nplus = \fG 0,
\label{NonlNeg}
\end{equation}
where $\fG 0 \in {\rm End}(\mythbf G)$ is the trivial endomorphism of
any element to unit
$\I \in \mythbf G$, $\fG 0(g) = \I$, $\forall g \in \mythbf G$. The endomorphism
should not be mixed with identity map $\fG 1(g) = g$.

Due to \eq{Rdistr} (left distributivity), \eq{Ldistr} (right distributivity)
and \eq{CommCond} here is some ``nonlinear analogue'' of {\em ring} with
zero is $\fG 0$, unit is $\fG 1$.

\subsubsection{Semidirect products, rings and algebras}

Let the group $\bf G$ has a commutative subgroup $\bf B$ and we are working
only with $\bf B$-invariant endomorphisms of $\bf G$. It is possible to consider
restrictions of such functions on $\bf B$, i.e. endomorphisms
of $\bf B$ ``recursively defined'' using expressions with ``constants'' from
group $\bf G$ (see examples below). But for endomorphisms of $\bf B$
\eq{Ldistr}, \eq{Rdistr} show distributivity of ``$\nplus$'' with respect
to composition. If $\fG{f,g}$ are endomorphisms, $\fG{f g}$ is also
endomorphism. For commutative group $\fG{f \nplus g = g \nplus f}$
is endomorphism due to \eq{CommCond}. So compositions of any given set
of endomorphisms $\bf B$ together with addition introduced by commutative
operation ``$\nplus$'' and negation \eq{NonlNeg} generate a ring of
endomorphisms.

\smallskip

If $\bf B$ is isomorphic with a vector space,\footnote{An abelian group may
also be not isomorphic with a vector space, for example $U(1)$ and torus, but
further in this paper are discussed only additive abelian groups isomorphic
with $\C^n$ or $\R^n$.} it is possible to introduce complete structure
of algebra on the ring by definition of action of scalars
$\fG f \mapsto \lambda \fG f$ on the space of endomorphisms $\bf B$ simply as:
\begin{equation}
 (\lambda \fG f)(b) \equiv \fG f(\lambda b); \quad
 \fG f, \lambda \fG f \in {\rm End}(\mythbf B).
\label{ScalRing}
\end{equation}

The action of scalars and ``$\nplus$'' satisfies to necessary identities:
\[%\begin{equation}
(\alpha \fG f \nplus \beta \fG f)(b) =
(\alpha \fG f)(b)\,(\beta \fG f)(b)  = \fG f(\alpha b)\,\fG f(\beta b)
\stackrel{_{\fG f \in {\rm End}(\mythbf B)}}{=} \fG f(\alpha b + \beta b)
%= \fG f\bigl((\alpha + \beta) b\bigr)
= \bigl((\alpha + \beta)\fG f\bigr) (b)
\]%\end{equation}
\[%\begin{equation}
(\alpha \fG f \nplus \alpha \fG g)(b) =
(\alpha \fG f)(b)\,(\alpha \fG g)(b)  = \fG f(\alpha b)\, \fG g(\alpha b)
= (\fG f \nplus \fG g)(\alpha b) = \bigl( \alpha (\fG f \nplus \fG g)\bigr)(b)
\]%\end{equation}

\medskip

It was already mentioned, that for the group $\mythbf{G = B \rtimes L}$
the additive abelian subgroup $\bf B$ can be considered as subspace of pairs
$(b,\I) \in \bf G$ and the subgroup $\bf L$ corresponds to $(\O,l) \in \bf G$.

If the additive abelian group $\bf B$ is isomorphic with vector space, the
endomorphisms --- are simply linear transformations (by definition).

A natural example of endomorphism of $\bf B$ is automorphism\footnote{
I.e. invertible endomorphism.}, expressed by inner automorphism of $\bf G$,
${\rm Ad}[g]\colon b \mapsto g b g^{-1}$.
For arbitrary semidirect product it is also possible to check:
\begin{equation}
 {\rm Ad}[(a,l)](b,\I) = (l(b),\I);
 \quad (a,l) \in \mythbf G;\ (b,\I),\,(l(b),\I) \in \mythbf{B \subset G}.
\label{AdB}
\end{equation}
It was already discussed
in relation with \eq{AdLP} and matrix representation \eq{AdMat}.
It is clear from \eq{AdB}, the restriction of the inner automorphism on $\bf B$
does not depends on $a$ and an element $(\O,l) \in \bf L$ may be used instead,
and if $\bf B$ is the vector space, then \eq{AdB} is simply a linear
transformation $b \mapsto lb$, $b \in \bf B$, $l \in \bf L$ similarly with
expression \eq{AdLP} for algebras used before.

Let us use shortenings:
\begin{equation}
 [l](b) \equiv {\rm Ad}[(\O,l)](b,\I) = (l(b),\I) \equiv l(b),\quad
 l \in \mythbf L,\ b \in \mythbf B,\ [l] \in {\rm End}(\mythbf B)
\end{equation}

It is possible to check, that the composition of such automorphisms
is simply $[l_1][l_2] = [l_1 l_2]$.
Now let us consider $[l_1] \nplus [l_2] \in {\rm End}(\mythbf B)$:
\begin{equation}
 ([l_1] \nplus [l_2])(b) = (l_1(b),\I) (l_2(b),\I) = (l_1(b) + l_2(b),\I)
 = l_1(b) + l_2(b).
\label{NonlAdd}
\end{equation}
It was not possible simply write $(l_1+l_2) b$ instead of
$l_1(b) + l_2(b)$, because for elements $l_1, l_2$ of the group $\bf L$
was defined only product, not addition, but anyway \eq{NonlAdd} defines some
commutative operation ``$\nplus$'' distributive with compositions
and we have a ring of endomorphisms.

If $\bf B$ is isomorphic with a vector space, then the ring is a subset of an
algebra
of linear operators on the space and it is possible to introduce a structure
of the algebra on the ring by definition action of scalars as in \eq{ScalRing}.
Let us denote such algebra $\nA\mythbf G$ or $\nA(\mythbf{B \rtimes L})$.

\smallskip

\noindent{\bf Note:}
It should be mentioned also, that if group $\bf L$ has some matrix (or
algebraic) representation, then it is possible to write $\hat l_1 + \hat l_2$
or $\lambda \hat l$, but anyway in general case
$[\hat l_1] \nplus [\hat l_2] \ne [\hat l_1 + \hat l_2]$ and even
$\lambda[\hat l] \ne [\lambda\hat l]$. An example is $\nA(\astar\D(\A))$
(relevant to spinor Poincar\'e group) there
$\lambda[\hat l] = [|\lambda|^2 \hat l]$.

\subsubsection{``Restoring'' of algebraic structures}

Let us consider particular cases of algebra $\nA$ for groups $\D(\A)$,
$\T(A)$ and $\astar\D(\A)$.

The $\D(\A)$ is represented as $\A \rtimes \Ginv\A$. Let us consider a pair
$(a,l) \in \D(\A)$, $a \in \A$, $l \in \Ginv\A$, because
$\Ginv\A \subset \A$, it is possible to consider $l$ as element of
algebra $\A$ and write $l_1 + l_2$ and $\lambda l$. On the other
hand, all invertible elements of $\A$ can be considered also as elements
of group $\Ginv\A$ and because action of element $l \in \Ginv\A$ on
$b \in \A$ is defined as left multiplication, it is simple to check for
$\D(\A)$ :
\begin{equation}
 \lambda[l] = [\lambda l], \quad \forall \lambda \ne 0,
\label{laml}
\end{equation}
and also from \eq{NonlAdd} for $\D(\A)$ follows:
\begin{equation}
 l_1 + l_2 \in \Ginv\A \Longrightarrow [l_1] \nplus [l_2] = [l_1 + l_2].
\label{addl}
\end{equation}
and so, {\em if span of all invertible elements $\Ginv\A$ is full algebra
$\A$, then $\nA\D(\A) \cong \A$}.

\medskip

Although $\astar\D(\A)$ is also may be represented as $\A \rtimes \Ginv\A$
the \eq{laml} here is not true, because action of an element $l \in \Ginv\A$
on $b \in \A$ is defined as $l\colon b \mapsto \hat{l} b \hat{l}^*$ and so
\begin{equation}
 \lambda[l](b) \equiv [l](\lambda b) = \hat{l} \lambda b \hat{l}^* \ne
 [\lambda l](b) = \lambda \hat{l} b \hat{l}^* \bar\lambda
\label{lamll}
\end{equation}
and \eq{addl} also is not true:
\begin{equation}
 ([l_1] \nplus [l_2])(b) = \hat l_1 b\,\hat l_1^* + \hat l_2 b\,\hat l_2^*
\ne [l_1 + l_2](b) = (\hat l_1 + \hat l_2)\,b\,(\hat l_1^* + \hat l_1^*)
\label{addll}
\end{equation}
and here really we see two different algebraic structures.
%\footnote{``classical'' and ``quantum''}

\smallskip

And, finally, for ``biggest'' group $\T(\A)$, $\D(\A) \subset \T(\A)$,
$\astar\D(\A) \subset \T(\A)$ it is possible to show, that {\em if span
of all invertible elements $\Ginv\A$ is full algebra $\A$, then
$\nA\T(\A) \cong \A \otimes \A$}.

It is clear also, that instead of $\Ginv\A$ in constructions
$\D(\A)$ or $\T(\A)$ it is possible to use any subgroup
$\Ginv\A' \subset \Ginv\A$ with property: ${\rm span}(\Ginv\A') = \A$
and for groups $\D'(\A)$, $\T'(\A)$ constructed by using $\Ginv\A'$ is
again true:
\begin{equation}
 \nA\D'(\A) \cong \A, \quad \nA\T'(\A) \cong \A \otimes \A.
\end{equation}

\subsubsection{Pauli algebra and complex Poincar\'e group}

Let us now consider the Pauli algebra $\PA$. A group of invertible elements
\mbox{$\Ginv\PA = GL(2,\C)$} and a subgroup $\Ginv\PA' \equiv SL(2,\C)$
satisfy the condition ${\rm span}(\Ginv\A) = {\rm span}(\Ginv\A') = \A$
and so structure of the Pauli algebra may be ``recovered'' using
an affine group of $\C^2$, i.e. $\D(\PA) = IGL(2,\C)$, or subgroup
$\D'(\PA) = ISL(2,\C)$:
$$\PA = \nA\D(\PA) = \nA\bigl(IGL(2,\C)\bigr)
 = \nA\D'(\PA) = \nA\bigl(ISL(2,\C)\bigr),$$
but the groups seem do not have some interesting applications in physics.

On the other hand, it was shown, that the spinor Poincar\'e group is a subgroup
of $\astar\D(\PA)$, but as it was discussed in relation with \eq{addll},
the algebra $\nA\bigl(\astar\D(\PA)\bigr)$ is not the Pauli algebra.%
%\footnote{But $\nA\astar\D(\PA)$ has close relation with
%mixed states and density matrices.}

\smallskip

So it is convenient to consider ``a complex conformal Poincar\'e
group'' $\tilde\T(\PA)$ \eq{tTPA} or its subgroup like $ISO(4,C)$.
The groups contain classical and spinor Poincar\'e group together
with affine group $ISL(2,\C)$ as subgroups.

\section{Conclusion}

For establishing examples of relations of the group of space-time symmetries
and algebraic structures on the space of quantum states here was used the
Pauli algebra, usually related with half-spinor representations of spin-1/2
particles. It was shown, that using simple constructions based on formal
treatment of algebras as groups with operators it is possible ``to recover''
structure of the Pauli algebra and a linear manifold of quantum
states from the group of symmetries of complex space-time, but application
of similar construction $(\nA)$ to the spinor Poincar\'e group generates another
algebra with additive structure \eq{addll}, more appropriate for manipulations
with density matrices (and probabilities).

On the other hand, complex Poincar\'e group $ISO(4,\C)$ includes classical
and spinor Poincar\'e groups, an affine group of a two-dimensional complex space
together with a conjugated group defined by \eq{starTA}, --- it is compatible
with structure of the $2D$ Hilbert space. Application of the discussed
constructions $(\nA)$ to complex Poincar\'e group ``recovers''
additive structure of the $2D$ Hilbert space of quantum states related
with composition of wave vectors, but also associated with structures induced
by its real spinor Poincar\'e subgroup discussed above.
%consistent with additivity of probabilities.

These last properties of complex Poincar\'e group may be interesting in
relation with questions of nonlocality in quantum mechanics --- really,
it just was mentioned that additive probabilistic picture compatible in
such approach with structure of (spinor) Poincar\'e group, but for
consistency with description of wave vectors and Hilbert spaces may be
relevant more wide group of symmetries, like complex Poincar\'e group.

A question about nonlinearities affected in Weinberg's paper \cite{wei}
was practically not discussed here, but possibly {\em a quasi-ring} discussed
in section \ref{part:nring} may be interesting from such point of view.
It should be mentioned also, that although operation ``$\nplus$'' defined
here coincides with addition for semidirect products like the Poincar\'e
group and defines a linear structure, from a ``recursive'' point of view it is
nonlinear operation, as it may be clear from definition \eq{npDef}. So despite
of ``reduction'' to the standard algebraic structure of quantum mechanics
for (complex) Poincar\'e group, the quasi-algebra $\nA$ may generate some
non-algebraic structures for more general groups like symmetry group $SO(4,1)$
of {\em de Sitter space}. On the other hand, even for this case (or for more
difficult models of general relativity) {\em locally} we have usual
Poincar\'e group and so there are no difficulties with introduction of
algebraic structures {\em locally} using the constructions like above.

{\def\newpage{\relax}

}
\end{document}